\documentclass{article}
\usepackage{cite}
\usepackage{graphicx}
\usepackage{dcolumn}
\usepackage{amsmath}

\begin{document}

\date{}
\title{On a class of non-Hermitian Hamiltonians with tridiagonal matrix
representation}
\author{Francisco M. Fern\'{a}ndez\thanks{%
fernande@quimica.unlp.edu.ar} \\
INIFTA, DQT, Sucursal 4, C. C. 16, \\
1900 La Plata, Argentina}
\maketitle

\begin{abstract}
We show that some non-Hermitian Hamiltonian operators with tridiagonal
matrix representation may be quasi Hermitian or similar to Hermitian
operators. In the class of Hamiltonian operators discussed here the
transformation is given by a Hermitian, positive-definite, diagonal
operator. We show that there is an important difference between open
boundary conditions and periodic ones. We illustrate the theoretical results
by means of two simple, widely used, models.
\end{abstract}

\section{Introduction}

\label{sec:intro}

Non-Hermitian quantum mechanics has become quite popular in recent years
because of its intrinsic mathematical interest and as a suitable tool for
the interpretation of some physical phenomena\cite{BBJ03,B07,M11} (and
references therein). Some non-Hermitian Hamiltonians exhibit PT-symmetry\cite
{BB98} that is a particular case of antiunitary symmetry\cite{W60} (some
simple examples are discussed elsewhere\cite{F16}).

In particular, exactly-solvable models given in terms of tridiagonal
matrices have proved useful for deriving and illustrating relevant
properties of non-Hermitian systems\cite
{Z07a,Z07b,Z07c,Z08a,Z08b,Z09,Z10,Z11}. Some non-Hermitian operators exhibit
generalized Hermiticity\cite{P43} or quasi Hermiticity\cite{SGH92} that
provides the condition for a linear operator to be similar to a self-adjoint
one\cite{W69}.

The purpose of this paper is a discussion of a class of non-Hermitian
operators with finite tridiagonal matrix representation somewhat more
general than those discussed earlier\cite
{Z07a,Z07b,Z07c,Z08a,Z08b,Z09,Z10,Z11}. In particular, we are interested in
simple versions of the Hatano-Nelson model\cite{HN96} that have recently
proved useful in the study of the effect of the boundary conditions and the
skin effect in non-Hermitian tight-binding models\cite{R21} and the
exceptional degeneracy and topological phases in such systems\cite{BBK21}.
More precisely, we study these models in the light of the above-mentioned
generalized Hermiticity\cite{P43} or quasi Hermticity\cite{SGH92} and the
conditions under which the non-Hermitian Hamiltonians are similar to
Hermitian ones\cite{W69}.

In section~\ref{sec:OBC} we derive the main results in the case of a
Hatano-Nelson model with open boundary conditions (OBC); in section~\ref
{sec:PBC} we apply those results to a similar model with periodic boundary
conditions (PBC); in section~\ref{sec:simple_model} we discuss
quasi-Hermiticity, Hermiticity and PT-symmetry by means of a simple model;
in section~\ref{sec:H-N model} we focus on the particular model discussed
briefly by Roccati\cite{R21} with both OBC and PBC; in section~\ref
{sec:Robust EP} we analyze a model proposed by Yuce and Ramezani\cite{YR19}
that illustrates the concept of robust exceptional points (EPs); finally, in
section~\ref{sec:conclusions} we summarize the main results of this paper
and draw conclusions.

\section{Open boundary conditions}

\label{sec:OBC}

We first consider a Hamiltonian operator with a tridiagonal matrix
representation and OBC
\begin{equation}
H=\sum_{j=1}^{N-1}\left( H_{j,j+1}\left| j\right\rangle \left\langle
j+1\right| +H_{j+1,j}\left| j+1\right\rangle \left\langle j\right| \right)
+\sum_{j=1}^{N}H_{j,j}\left| j\right\rangle \left\langle j\right| ,
\label{eq:H_OBC}
\end{equation}
where $H_{j,j}^{*}=H_{j,j}$. By means of a diagonal operator
\begin{equation}
Q=\sum_{j=1}^{N}Q_{j}\left| j\right\rangle \left\langle j\right|
,\;Q_{k}\neq 0,\;k=1,2,\ldots ,N,  \label{eq:Q}
\end{equation}
we carry out the transformation
\begin{eqnarray}
\tilde{H} &=&Q^{-1}HQ=\sum_{j=1}^{N-1}\left( \tilde{H}_{j,j+1}\left|
j\right\rangle \left\langle j+1\right| +\tilde{H}_{j+1,j}\left|
j+1\right\rangle \left\langle j\right| \right)  \nonumber \\
&&+\sum_{j=1}^{N}H_{j,j}\left| j\right\rangle \left\langle j\right| ,
\nonumber \\
\tilde{H}_{j,j+1} &=&\frac{Q_{j+1}}{Q_{j}}H_{j,j+1},\;\tilde{H}_{j+1,j}=%
\frac{Q_{j}}{Q_{j+1}}H_{j+1,j}.  \label{eq:H_transf}
\end{eqnarray}
If we require that $\tilde{H}^{\dagger }=\tilde{H}$ then
\begin{equation}
\frac{H_{j+1,j}^{*}}{H_{j,j+1}}=\left| \frac{Q_{j+1}}{Q_{j}}\right|
^{2}=R_{j}>0.  \label{eq:R_j}
\end{equation}
This equation is a sufficient condition for $H$ to be similar to a Hermitian
operator. Note that if $R_{j}=1$ for all $j$ the operator $H$ is Hermitian.
It is clear that if the nonzero off-diagonal matrix elements are of the form
$H_{j,j+1}=r_{j}e^{i\theta _{j}}$, $H_{j+1,j}=\rho _{j}e^{-i\theta _{j}}$, $%
r_j ,\rho_j > 0$, then the non-Hermitian Hamiltonian operator (\ref{eq:H_OBC}%
) is similar to a Hermitian one and $R_{j}=\rho _{j}/r_{j}$.

According to equation (\ref{eq:R_j}) it is sufficient for present purposes
to choose $Q_{j}=\left| Q_{j}\right| $ so that $Q$ is both Hermitian ($%
Q^{\dagger }=Q$) and positive definite ($\left\langle Q\right\rangle >0$).
It follows from $\tilde{H}^{\dagger }=\tilde{H}$ that $Q^{2}H^{\dagger
}=HQ^{2}$ which is the condition required by Scholtz et al\cite{SGH92} for
quasi Hermiticity and also by Williams's theorems\cite{W69}. Besides, it
follows from equation (\ref{eq:R_j}) that
\begin{equation}
Q_{j}=\sqrt{R_{j-1}R_{j-2}\ldots R_{1}}Q_{1}.  \label{eq:Q_j...Q_1}
\end{equation}
Since $\left( Q_{1}^{-1}Q\right) ^{-1}H\left( Q_{1}^{-1}Q\right) =Q^{-1}HQ$
we can choose $Q_{1}=1$ without loss of generality.

The eigenvalue equation $H\psi _{n}=E_{n}\psi _{n}$ becomes $\tilde{H}%
Q^{-1}\psi _{n}=E_{n}Q^{-1}\psi _{n}$ under the transformation (\ref
{eq:H_transf}). The eigenvectors $\tilde{\psi}_{n}=Q^{-1}\psi _{n}$ of $%
\tilde{H}$ can be chosen to be orthonormal; therefore, $\left\langle \tilde{%
\psi}_{m}\right| \left. \tilde{\psi}_{n}\right\rangle =\left\langle \psi
_{m}\right| Q^{-2}\left| \psi _{n}\right\rangle =\delta _{mn}$ suggests that
we can choose the metric $\eta =Q^{-2}$ as argued by Pauli\cite{P43} several
years ago.

Since the dimensionless time-dependent equation\cite{F20}
\begin{equation}
i\frac{\partial }{\partial t}\psi (t)=H\psi (t),  \label{eq:Schro_td}
\end{equation}
can be transformed into
\begin{equation}
i\frac{\partial }{\partial t}Q^{-1}\psi (t)=\tilde{H}Q^{-1}\psi (t),
\label{eq:Schro_td_2}
\end{equation}
then its solution is given by
\begin{equation}
Q^{-1}\psi (t)=\exp \left( -it\tilde{H}\right) Q^{-1}\psi (0),
\label{eq:Schro_td_sol}
\end{equation}
where the time-evolution operator $\exp \left( -it\tilde{H}\right) $ is
obviously unitary.

\section{Periodic boundary conditions}

\label{sec:PBC}

In the case of PBC ($\left| N+1\right\rangle =\left| 1\right\rangle $) we
write $H_{N+1,N}=H_{1,N}$ and $H_{N,N+1}=H_{N,1}$ so that the Hamiltonian
operator reads
\begin{eqnarray}
H &=&\sum_{j=1}^{N-1}\left( H_{j,j+1}\left| j\right\rangle \left\langle
j+1\right| +H_{j+1,j}\left| j+1\right\rangle \left\langle j\right| \right)
\nonumber \\
&&+H_{1,N}\left| 1\right\rangle \left\langle N\right| +H_{N,1}\left|
N\right\rangle \left\langle 1\right| +\sum_{j=1}^{N}H_{j,j}\left|
j\right\rangle \left\langle j\right| .  \label{eq:H_PBC}
\end{eqnarray}
In addition to the expressions given in equation (\ref{eq:H_transf}) we have
\begin{eqnarray}
\tilde{H}_{1,N} &=&\frac{Q_{N}}{Q_{1}}H_{1,N}=\sqrt{R_{N-1}R_{N-2}\ldots
R_{1}}H_{1,N},  \nonumber \\
\tilde{H}_{N,1} &=&\frac{Q_{1}}{Q_{N}}H_{N,1}=\frac{H_{N,1}}{\sqrt{%
R_{N-1}R_{N-2}\ldots R_{1}}},  \label{eq:H_N1_transf_PBC}
\end{eqnarray}
and the transformed Hamiltonian operator $\tilde{H}$ is not Hermitian unless
$H_{1.N}$ and $H_{N,1}$ are chosen conveniently. The additional requirement $%
\tilde{H}_{N,1}=\tilde{H}_{1,N}^{*}$ leads to
\begin{equation}
H_{N,1}=R_{N-1}R_{N-2}\ldots R_{1}H_{1,N}^{*}.  \label{eq:H_N1_PBC}
\end{equation}
Also in this case we have that $H_{N,1}=r_{N}e^{i\theta _{N}}$ and $%
H_{1,N}=\rho _{N}e^{-i\theta _{N}}$ but the ratio $\rho _{N}/r_{N}$ depends
on all the other ratios $R_{j}$. If we define $R_{N}=H_{1,N}^{*}/H_{N,1}$
then $R_{N}R_{N-1}\ldots R_{1}=1$, where all the $R_{j}$, $j=1,2,\ldots ,N-1$%
, are arbitrary positive numbers.

Note that if we set $H_{1,N}=0$ in equation (\ref{eq:H_N1_PBC}) then $%
H_{N,1}=0$ and we recover the open chain.

\section{Simple model}

\label{sec:simple_model}

In order to illustrate that there is no relevant connection between
quasi-Hermiticity\cite{SGH92} and PT-symmetry\cite{BB98} (or, more
generally, antiunitary symmetry\cite{W60}), we choose the simplest
non-Hermitian model given by
\begin{equation}
\mathbf{H}=\left(
\begin{array}{ll}
a & re^{i\theta } \\
\rho e^{-i\theta } & b
\end{array}
\right) ,  \label{eq:H_2x2_QH}
\end{equation}
where $a$, $b$ and $\theta $ are real and $r,\rho >0$. This matrix is
neither Hermitian or PT-symmetric but it is similar to the Hermitian one
\begin{equation}
\mathbf{\tilde{H}}=\left(
\begin{array}{ll}
a & \sqrt{r\rho }e^{i\theta } \\
\sqrt{r\rho }e^{-i\theta } & b
\end{array}
\right) ,  \label{eq:H_2x2_transf}
\end{equation}
through the transformation given by
\begin{equation}
\mathbf{Q}=Q_{1}\left(
\begin{array}{ll}
1 & 0 \\
0 & \sqrt{\frac{\rho }{r}}
\end{array}
\right) .  \label{eq:Q_2x2}
\end{equation}

If $r=\rho $ and $a\neq b$ then the matrix (\ref{eq:H_2x2_QH}) is not
PT-symmetric but it is Hermitian, in which case $\mathbf{Q}=Q_{1}\mathbf{I}$%
, where $\mathbf{I}$ is the $2\times 2$ identity matrix.

If $a=b$ and $r=\rho $ the matrix (\ref{eq:H_2x2_QH}) is PT-symmetric with
the antiunitary symmetry based on the orthogonal matrix (parity)
\begin{equation}
\mathbf{P}=\left(
\begin{array}{ll}
0 & 1 \\
1 & 0
\end{array}
\right) .  \label{eq:P_2x2}
\end{equation}
However, this particular case is also Hermitian.

Finally, the matrix
\begin{equation}
\mathbf{H}^{PT}=\left(
\begin{array}{ll}
\rho e^{i\beta } & re^{i\theta } \\
re^{-i\theta } & \rho e^{-i\beta }
\end{array}
\right) ,  \label{eq:H_2x2_PT}
\end{equation}
is neither Hermitian or quasi-Hermitian (in the sense discussed above) but
it is PT-symmetric with parity given by (\ref{eq:P_2x2}). In this case,
parity is broken when $r^{2}<\rho ^{2}\sin ^{2}\beta $. This model is
discussed in Appendix~\ref{sec:appendix} by means of a somewhat more general
approach.

\section{Simplified Hatano-Nelson model}

\label{sec:H-N model}

In this section we apply the results developed above to a model outlined by
Roccati\cite{R21} that is a simple version of the Hatano-Nelson model
without disorder\cite{HN96}. For clarity, we discuss the OBC and PBC
separately in two short subsections.

\subsection{Example with OBC}

\label{subsec:Example_OBC}

In this example $H_{j,j}=0$, $H_{j,j+1}=J(1-\delta )$, $H_{j+1,j}=J(1+\delta
)$, where $J$ and $\delta $ are real model parameters. The condition $%
H_{j+1,j}^{*}/H_{j,j+1}=R_{j}=(1+\delta )/(1-\delta )>0$ leads to $-1<\delta
<1$ in agreement with Roccati's choice\cite{R21}. The tridiagonal matrix
representation of the transformed Hamiltonian $\tilde{H}$ is symmetric with $%
\tilde{H}_{j,j+1}=\tilde{H}_{j+1,j}=J\sqrt{1-\delta ^{2}}$. We conclude that
all the eigenvalues are proportional to $J\sqrt{1-\delta ^{2}}$.

\subsection{Example with PBC}

\label{subsec:example_PBC}

Roccati\cite{R21} considered the PBC explicitly. If we take into account
that $H_{N,N+1}=H_{N,1}=J(1-\delta )$ and $H_{N+1,N}=H_{1,N}=J(1+\delta )$
then the resulting non-Hermitian Hamiltonian exhibits real and complex
eigenvalues. However, if we choose $H_{1,N}$ as before and
\begin{equation}
H_{N,1}=J\frac{(1+\delta )^{N}}{(1-\delta )^{N-1}},  \label{eq:H_N!_PBC}
\end{equation}
then the eigenvalues are real because
\begin{equation}
\tilde{H}_{N,1}=\tilde{H}_{1,N}=\frac{(1+\delta )^{\frac{N+1}{2}}}{(1-\delta
)^{\frac{N-1}{2}}}.  \label{eq:H_N1=H_1N_transf_PBC}
\end{equation}

\begin{figure}[tbp]
\begin{center}
\includegraphics[width=9cm]{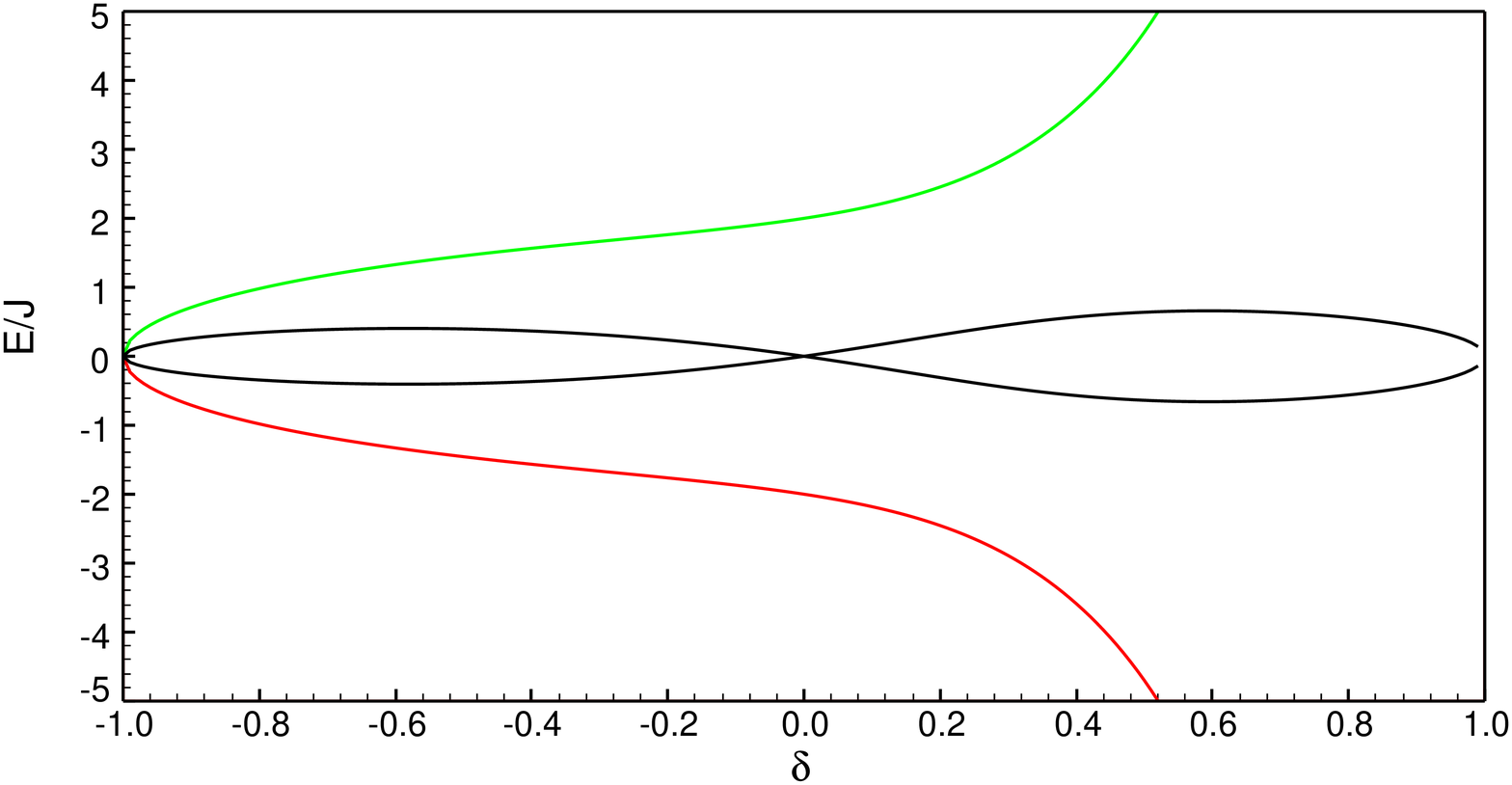}
\end{center}
\caption{Eigenvalues $E_n/J$ of the Roccati's model\protect\cite{R21} with
PBC and $N=4$}
\label{Fig:H4PBO}
\end{figure}

Figure~\ref{Fig:H4PBO} shows the eigenvalues $E_{n}/J$ for this model when $%
N=4$ as functions of $\delta $. There is a level crossing at $\delta =0$
because of the higher symmetry of the resulting Hermitian operator with all
nonzero matrix elements equal to $J$ (it resembles a H\"{u}ckel matrix\cite
{P68}).

The behaviour of the eigenvalues at $\delta =-1$ is
\begin{eqnarray}
E_{1} &=&-E_{4}\sim -\left( \frac{\sqrt{10}}{2}+\frac{\sqrt{2}}{2}\right)
\sqrt{\sigma }+\mathcal{O}\left( \sigma ^{3/2}\right) ,  \nonumber \\
E_{2} &=&-E_{3}\sim \left( \frac{\sqrt{2}}{2}-\frac{\sqrt{10}}{2}\right)
\sqrt{\sigma }+\mathcal{O}\left( \sigma ^{3/2}\right) ,\;\sigma =\delta +1.
\label{eq:E_n_left}
\end{eqnarray}
On the other hand, at $\delta =1$ we have
\begin{eqnarray}
E_{1} &=&-E_{4}\sim -4\,{\xi }^{-1}+4-\xi -{\frac{1}{2}}{\xi }^{2}-{\frac{3}{%
8}}{\xi }^{3}+\mathcal{O}\left( {\xi }^{4}\right) ,  \nonumber \\
E_{2} &=&-E_{3}\sim -\sqrt{2\xi }+\,\frac{{\xi }^{3/2}}{\sqrt{2}}+\mathcal{O}%
\left( {\xi }^{5/2}\right) ,\;\xi =1-\delta .  \label{eq:E_n_right}
\end{eqnarray}
We appreciate that there is a pole at $\delta =1$ that comes from the pole
in the matrix element $H_{N,1}$.

\begin{figure}[tbp]
\begin{center}
\includegraphics[width=9cm]{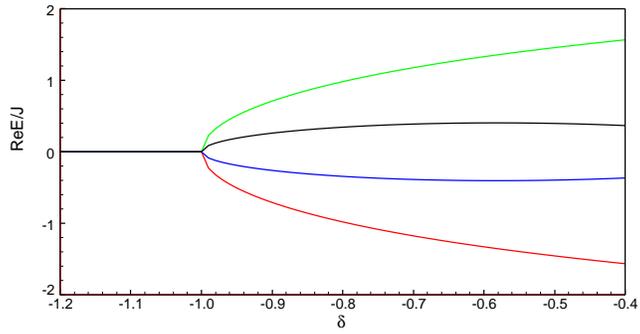}
\end{center}
\caption{Real part of the eigenvalues $E_n/J$ of the Roccati's model%
\protect\cite{R21} with PBC and $N=4$.}
\label{Fig:H4PBOR}
\end{figure}

\begin{figure}[tbp]
\begin{center}
\includegraphics[width=9cm]{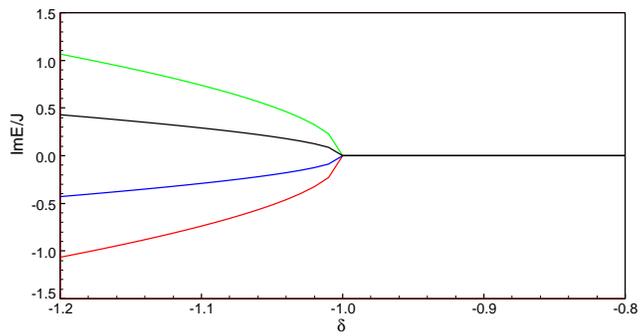}
\end{center}
\caption{Imaginary part of the eigenvalues $E_n/J$ of the Roccati's model%
\protect\cite{R21} with PBC and $N=4$.}
\label{Fig:H4PBOI}
\end{figure}

At $\delta =-1$ there is an EP of order four as shown in figures \ref
{Fig:H4PBOR} and \ref{Fig:H4PBOI} for the real and imaginary parts of $%
E_{n}/J$, respectively. In fact, when $\delta =-1$ the $4\times 4$
Hamiltonian matrix exhibits just one eigenvector.

\section{Robust exceptional point}

\label{sec:Robust EP}

If an EP does not change much with respect to perturbations of the system,
one commonly says that it is robust. In order to obtain robust EPs, Yuce and
Ramezani (YR)\cite{YR19} proposed a non-Hermitian tight binding Hamiltonian
that is a particular case of
\begin{equation}
H=\sum_{j=1}^{N-1}\left( t_{n}\left| n\right\rangle \left\langle n+1\right|
+J_{n}\left| n+1\right\rangle \left\langle n\right| \right)
+\sum_{j=1}^{N}\beta _{n}\left| n\right\rangle \left\langle n\right| ,
\label{eq:H_Yuce}
\end{equation}
where $\beta _{n}$, $t_{n}$ and $J_{n}$ are real model parameters, $t_{n}$
and $J_{n}$ being the forward and backward tunnelings, respectively. If $%
t_{n}J_{n}>0$ for all $n$, this operator satisfies the conditions outlined
in section~\ref{sec:OBC} and, consequently, is similar to a Hermitian one.
YR considered the particular case given by $\beta _{n}=\beta $ and $%
J_{n}=t_{n}-\gamma \delta _{n,N-1}$, $n=1,2,\ldots ,N-1$, with $t_{n}>0$.
Clearly, this operator is quasi-Hermitian and isospectral to a Hermitian one
provided that $t_{N-1}\left( t_{N-1}-\gamma \right) >0$, disregarding the
values of the other model parameters because $t_{n}J_{n}=t_{n}^{2}>0$, $%
n=1,2,\ldots ,N-2$. Since $t_{N-1}>0$, then the exceptional point $\gamma
_{EP}$ should appear at $\gamma _{EP}\leq t_{N-1}$ because the condition
just mentioned is sufficient but not necessary. Present argument shows that
the random change of the parameters $t_{n}$ (the disorder) will not affect
the location of the EP indicated above as long as $t_{N-1}$ is kept
unchanged. We can easily generalize YR's results. If we keep $t_{k}>0$
unchanged and define $J_{n}=t_{n}-\gamma \delta _{nk}$, then the arbitrary
variation of the real model parameters $t_{n}$, $n\neq k$, will not change
the fact that $\gamma _{EP}\leq t_{k}$. Obviously, if $t_{k}<0$ then $\gamma
_{EP}\geq t_{k}$. It is clear that the location of the EP is independent of $%
\beta $ (it is only necessary to define $\epsilon =E-\beta $ to realize that
we may choose $\beta =0$ without loss of generality).

In order to determine $\gamma _{EP}$ we resort to the discriminant of the
characteristic polynomial
\begin{equation}
p_{N}(E,\gamma )=\det \left( \mathbf{H}_{N}-E\mathbf{I}_{N}\right) ,
\label{eq:charpoly}
\end{equation}
where $\mathbf{H}_{N}$ and $\mathbf{I}_{N}$ are the tridiagonal matrix
representation of $H$ and the $N\times N$ identity matrix, respectively. The
discriminant $F_{N}(\gamma )=Disc_{E}\left( p_{N}(E,\gamma )\right) $ (see,
for example, \cite{AF21} and references therein) is a polynomial function of
$\gamma $ of degree $2N-3$ and the EPs are roots of $F_{N}(\gamma )$.

For one of YR's particular cases: $t_{n}=1$, $n=1,2,\ldots ,N-1$, the
argument above predicts that $\gamma _{EP}\geq 1$. The following results:
\begin{eqnarray}
F_{2} &=&4(1-\gamma ),  \nonumber \\
F_{3} &=&4(2-\gamma )^{3},  \nonumber \\
F_{4} &=&16\left( 1-\gamma \right) \left( \gamma ^{2}-2\gamma +5\right) ^{2},
\nonumber \\
F_{5} &=&16\left( 3-2\gamma \right) ^{3}\left( \gamma ^{2}+4\right) ^{2},
\nonumber \\
F_{6} &=&64\left( 1-\gamma \right) \left( 5\gamma ^{4}-8\gamma ^{3}+18\gamma
^{2}-28\gamma +49\right) ^{2},  \nonumber \\
F_{7} &=&64\left( 4-3\gamma \right) ^{3}\left( 4\gamma ^{4}+13\gamma
^{2}+32\right) ^{2},  \nonumber \\
F_{8} &=&256\left( 1-\gamma \right) \left( 49\gamma ^{6}-70\gamma
^{5}+151\gamma ^{4}-212\gamma ^{3}+351\gamma ^{2}-486\gamma +729\right) ^{2},
\nonumber \\
F_{9} &=&256\left( 5-4\gamma \right) ^{3}\left( 32\gamma ^{6}+93\gamma
^{4}+204\gamma ^{2}+400\right) ^{2},  \nonumber \\
F_{10} &=&1024\left( 1-\gamma \right) \times \left( 729\gamma ^{8}-972\gamma
^{7}+2052\gamma ^{6}-2704\gamma ^{5}\right.  \nonumber \\
&&\left. +4350\gamma ^{4}-5676\gamma ^{3}+8228\gamma ^{2}-10648\gamma
+14641\right) ^{2},  \nonumber \\
F_{11} &=&1024\left( 6-5\gamma \right) ^{3}\left( 400\gamma ^{8}+1084\gamma
^{6}+2213\gamma ^{4}+4032\gamma ^{2}+6912\right) ^{2},  \label{eq:F_N}
\end{eqnarray}
suggest that $\gamma _{EP}=1$ when $N=2K$, $K=1,2,\ldots $, and $\gamma
_{EP}=(K+1)/K$ when $N=2K+1$, in agreement with the theoretical prediction.
We realize that $\gamma _{EP}=1$ for all even values of $N$ but it changes
when $N$ is odd in such a way that it approaches unity as $N$ increases.
Note the multiplicity of the EPs for $N$ odd; for example, $%
p_{5}(E,3/2)=E^{3}\left( 5-2E^{2}\right) /2$ reveals an EP of order three
(there is only one eigenvector with eigenvalue $E=0$).

As a second example, we generated $t_{n}$, $n=1,2,\ldots ,N-1$, and $J_{n}$,
$n=1,2,\ldots ,N-2$ randomly in the interval $(0,1)$ and kept $%
J_{N-1}=1-\gamma $ fixed. The results suggest that $\gamma _{EP}=1$ for all $%
N=2K$ while the value of $\gamma _{EP}$ changes, but remains greater than
unity, for $N=2K+1$.

From numerical results for $\beta =2$ and $N=10$ YR\cite{YR19} concluded
that the EP at $\gamma =1$ is robust. Present theoretical results and
analytical calculations confirm this result that appears to be valid only
for even values of $N$. Apparently, YR did not try odd values of $N$ and
they even stated that ``Without loss of generality we assume that $N$ is an
even number.'' Another YR's wrong statement is ``The Hamiltonian in eq. (4)
becomes noninvertible when $\beta =0$.'' that only holds for odd values of $%
N $ because there is always an eigenvalue $E=\beta $. For even values of $N$%
, on the other hand, the Hamiltonian operator just mentioned is invertible
for $\beta =0$, except for some particular values of $\gamma $.

\section{Conclusions}

\label{sec:conclusions}

Several authors have discussed features of non-Hermitian quantum mechanics
by means of Hamiltonians with tridiagonal matrix representations\cite
{Z07a,Z07b,Z07c,Z08a,Z08b,Z09,Z10,Z11,R21,BBK21,YR19} (and references
therein). In many of those cases the occurrence of real eigenvalues can be
explained by the fact that the non-Hermitian Hamiltonians are quasi
Hermitian or similar to Hermitian ones. Throughout this paper we have
explored a class of such examples where the suitable transformation is given
by a Hermitian, positive-definite, diagonal operator. Quasi Hermiticity
appears more straightforwardly in the case of OBC, whereas for PBC one has
to choose a pair of matrix elements with somewhat more care. Section~\ref
{sec:Robust EP} shows how present theoretical result may clarify, explain
and even correct the conclusions drawn from numerical results on simple
lattice models.

\appendix

\numberwithin{equation}{section}

\section{More general approach}

\label{sec:appendix}

In this Appendix we will show that the matrix (\ref{eq:H_2x2_PT}) is also
similar to an Hermitian one. To this end, we resort to the more general
results of Williams\cite{W69} and Scholtz et al\cite{SGH92}.

Suppose that $H$ is a non-Hermitian operator that satisfies
\begin{equation}
H^{\dagger }G=GH,  \label{eq:HG=GH}
\end{equation}
where $G$ is an Hermitian, positive-definite operator. Then, $G^{-1}$ and $%
G^{1/2}$ exist and equation (\ref{eq:HG=GH}) can be rewritten as $%
G^{-1/2}H^{\dagger }G^{1/2}=G^{1/2}HG^{-1/2}$. Consequently,
\begin{equation}
\tilde{H}=G^{1/2}HG^{-1/2},  \label{eq:H^tilde}
\end{equation}
is Hermitian.

As an example we consider the simple PT-symmetric $2\times 2$ matrix
representation
\begin{equation}
\mathbf{H}=\left(
\begin{array}{ll}
a & b \\
b^{*} & a^{*}
\end{array}
\right) ,  \label{eq:H_2x2_PT_2}
\end{equation}
that is identical to (\ref{eq:H_2x2_PT}). In this case the matrix $\mathbf{G}
$ should be of the form $\mathbf{G}=\left(
\begin{array}{ll}
G_{11} & G_{12} \\
G_{12}^{*} & G_{22}
\end{array}
\right) $, where $G_{11}$ and $G_{22}$ are real. It follows from equation (%
\ref{eq:HG=GH}) that $G_{11}=G_{22}$ and $G_{11}\left( a^{*}-a\right)
-G_{12}b^{*}+bG_{12}^{*}=0$. Besides, it is clear from equation (\ref
{eq:HG=GH}) that we can choose $G_{11}=1$ without loss of generality;
therefore,
\begin{equation}
\mathbf{G}=\left(
\begin{array}{ll}
1 & G_{12} \\
G_{12}^{*} & 1
\end{array}
\right) ,  \label{eq:G}
\end{equation}
with the condition that
\begin{equation}
a^{*}-a-G_{12}b^{*}+bG_{12}^{*}=0.  \label{eq:G_condition}
\end{equation}
Since the eigenvalues of $\mathbf{G}$ are $g_{1}=1-\left| G_{12}\right| $
and $g_{2}=1+\left| G_{12}\right| $ we conclude that $\mathbf{G}$ is
positive-definite provided that $\left| G_{12}\right| <1$.

As a particular example we consider
\begin{equation}
\mathbf{H}=\left(
\begin{array}{ll}
i\gamma  & 1 \\
1 & -i\gamma
\end{array}
\right) .  \label{eq:H_2x2_PT_3}
\end{equation}
According to equation (\ref{eq:G_condition}) $G_{12}=-i\gamma $ and
\begin{equation}
\mathbf{G}=\left(
\begin{array}{ll}
1 & -i\gamma  \\
i\gamma  & 1
\end{array}
\right) .  \label{eq:G_2}
\end{equation}
Note that $\mathbf{G}$ is positive-definite provided that $\left| \gamma
\right| <1$, which is consistent with the fact that the eigenvalues of $%
\mathbf{H}$, $E_{1}=-\sqrt{1-\gamma ^{2}}$ and $E_{2}=-\sqrt{1-\gamma ^{2}}$%
, are real under the same condition. By means of
\begin{equation}
\mathbf{G}^{1/2}=\frac{1}{2}\left(
\begin{array}{ll}
\sqrt{1-\gamma }+\sqrt{\gamma +1} & i\left( \sqrt{1-\gamma }-\sqrt{\gamma +1}%
\right)  \\
-i\left( \sqrt{1-\gamma }-\sqrt{\gamma +1}\right)  & \sqrt{1-\gamma }+\sqrt{%
\gamma +1}
\end{array}
\right) ,  \label{eq:G_2^(1/2)}
\end{equation}
we obtain
\begin{equation}
\mathbf{\tilde{H}}=\left(
\begin{array}{ll}
0 & \sqrt{1-\gamma ^{2}} \\
\sqrt{1-\gamma ^{2}} & 0
\end{array}
\right) ,  \label{eq:H2x2_PT_3^tilde}
\end{equation}
that is a real symmetric matrix when $\left| \gamma \right| <1$.

The straightforward approach followed here appears to be simpler than the
one based on the left and right eigenvectors of $\mathbf{H}$\cite{SMM22}.

\end{document}